# From Disorder to Function: Entropy-Engineered Broadband Photonics with Ion-Transport-Stabilized Spectral Fidelity

Yuxiang Xin[†1, 2], Chen-Xin Yu[†1], Jianru Wang[1, 2], Jianbo Jin[3], Minliang Lai[3,4], Yinan Wang[5], Shuwen Yan[6], Gu-wen Chen[7], Liang Fan[1, 2], Xiachu Xiao[1, 2], Yutao Yang[1, 2], Luying Li[6], Han Wang[8,9], Zhi-Pan Liu[7, 10], Jiang Tang[6], Li-Ming Yang[*1] and Zhuolei Zhang[*1, 2]

[1]Y. Xin, C. Yu, J. Wang, L. Fan, X., Xiao, Y., Yang, L. M. Yang, Z. Zhang
School of Chemistry and Chemical Engineering, Hubei Key Laboratory of Bioinorganic Chemistry and Materia Medica, Hubei Engineering Research Center for Biomaterials and Medical Protective Materials, Key Laboratory of Material Chemistry for Energy Conversion and Storage, Ministry of Education, Huazhong University of Science and Technology (HUST), Wuhan 430074, P. R. China
E-mail: zzhuolei@hust.edu.cn
E-mail: Lmyang@hust.edu.cn

[2]Y. Xin, J. Wang, L. Fan, X., Xiao, Y., Yang, Z. Zhang
Guangdong HUST (Huazhong University of Science and Technology) Industrial Technology Research Institute Dongguan 523808, P.R. China

[3]J. Jin, M. Lai
[3]State Key Laboratory of Bioinspired Interfacial Materials Science, Suzhou Institute for Advanced Research, University of Science and Technology of China, Suzhou 215123, China

[4]M. Lai
[4]Department of Applied Chemistry, School of Chemistry and Materials Science, University of Science and Technology of China, Hefei 230026, China

[5]Y. Wang
Artificial Intelligence for Science Institute, No.150 Chengfu Road, Haidian District, Beijing, 100084, China

[6]S. Yan, L. Li, J. Tang
Wuhan National Laboratory for Optoelectronics (WNLO) and School of Optical and Electronic Information, Huazhong University of Science and Technology (HUST), 1037 Luoyu Road, Wuhan, Hubei 430074, China

[7]G. Chen, Z. P. Liu
State Key Laboratory of Porous Materials for Separation and Conversion, Collaborative Innovation Center of Chemistry for Energy Material, Shanghai Key Laboratory of Molecular Catalysis and Innovative Materials, Key Laboratory of Computational Physical Science, Department of Chemistry, Fudan University, Shanghai 200433, China

[8]H. Wang
National Key Laboratory of Computational Physics, Institute of Applied Physics and Computational Mathematics, Fenghao East Road 2, Beijing 100094, P.R. China

[9]H. Wang
HEDPS, CAPT, College of Engineering, Peking University, Beijing 100871, P.R. China

[10]Z. P. Liu
State Key Laboratory of Metal Organic Chemistry, Shanghai Institute of Organic Chemistry, Chinese Academy of Sciences, Shanghai 200032, China

[†] These authors contributed equally to this work.

## Abstract

The high-entropy halide-perovskite field has expanded rapidly, yet a key gap remains: configurational entropy is not yet a reliable, designable lever to co-deliver expanded photonic functionality and operational robustness with a composition-transferable mechanistic basis. Here we develop entropy-engineered rare-earth halide double-perovskite single crystals, $Cs_2Na(Sb, RE)Cl_6$ ($RE^{3+}$ = $Sc^{3+}$, $Er^{3+}$, $Yb^{3+}$, $Tm^{3+}$), that simultaneously expand near-infrared (NIR) functionality and establish a mechanistic stability rule. Near-equiatomic B(III)-site alloying yields a single-phase high-entropy solid solution ($\Delta S_{config} \approx 1.6R$). $Sb^{3+}$ serves as a sensitizer that unifies excitation and cooperatively activates multiple lanthanide channels, transforming the parent single-mode response into a broadband NIR output (~850–1600 nm) with three spectrally orthogonal fingerprint bands at 996, 1220 and 1540 nm. This tri-peak, self-referenced output enables redundancy-based ratiometric solvent identification and quantitative mixture sensing with reduced susceptibility to intensity drift. Accelerated aging under humidity and oxygen shows improved phase and emission stability versus single-component analogues. DFT and molecular dynamics attribute the robustness to strongly suppressed $RE^{3+}/Cl^{-}$ self-diffusion despite comparable $H_2O/O_2$ adsorption, kinetically impeding ion-migration-assisted reconstruction and degradation. Integration into a phosphor-converted LED delivers spectrally stable, broadband NIR illumination, establishing entropy engineering as a practical handle to couple expanded photonic functionality with mechanistically accountable durability in metal-halide photonics.

## Introduction

High-entropy materials, commonly defined as solids incorporating five or more elements in near-equiatomic ratios, offer a powerful thermodynamic lever to stabilize otherwise metastable lattices by increasing the entropic contribution ($-T\Delta S_{config}$) to the Gibbs free energy ($\Delta G$).[1, 2] This principle has reshaped materials design across alloys and ceramics and has recently entered the field of halide perovskites—an optoelectronic materials family whose operational durability is frequently limited by structural softness, interfacial reactivity, and fast ionic motion. Building on pioneering demonstrations of entropy-stabilized halide perovskite single crystals enabled by mild chemistry (e.g., Yang and co-workers)[3] and subsequent extensions to diverse high-entropy perovskite compositions,[4-6] the compositional space has expanded rapidly. Nevertheless, progress remains largely empirical: enhanced phase persistence is often observed, but whether configurational entropy can be used as a rational design variable for device-relevant photonics, and whether its stability benefit can be translated into mechanistic, transferable rules, remain open questions.[7-11]

A first gap concerns the functional consequences and opportunities of entropy-induced disorder. For photonic solids, compositional complexity is intrinsically ambivalent.[12] Local distortion and broadened microenvironments can suppress catastrophic phase separation, yet the same disorder can introduce nonradiative channels, amplify inhomogeneous fields, and destabilize the kinetic balance required for predictable optical operation.[13, 14] As a result, most high-entropy halide-perovskite emitters still behave effectively as single-channel systems, where disorder is largely “averaged” into a broadened response rather than translated into new device-relevant functions.[4, 5, 15] This is conceptually unsatisfying: high-entropy design deliberately incorporates multiple elements into one lattice, but their element-specific optical manifolds are seldom exploited as functional degrees of freedom. What is still missing is a general strategy that assigns complementary photonic roles (absorption/sensitization, energy routing, and emission) to different constituents, turning compositional disorder into a programmable handle for intrinsic photonic function expansion in a single crystalline emitter, instead of simply averaging it into spectral broadening, while maintaining high spectral fidelity and robustness.

A second gap concerns mechanistic attribution and transferability. In humid and oxidative environments, the degradation of halide perovskites is rarely explained by bulk thermodynamics alone;[16-18] instead, it frequently involves coupled interfacial and transport-mediated processes, where adsorption-triggered perturbations at surfaces/interfaces can propagate via ion migration and compositional redistribution, culminating in reconstruction and phase evolution accompanied by intensity and spectral drift.[19, 20] While high-entropy approaches are increasingly used to enhance durability, the origin of this apparent robustness remains poorly resolved, and it is still unclear which processes ultimately govern degradation across compositions and operating conditions. Developing ways to reveal these contributions with quantitative descriptors is therefore essential for moving from phenomenology to predictive design. Without explicitly separating these contributions and establishing quantifiable criteria, stability in high-entropy perovskites is unlikely to progress from empirical correlations to predictive design.

In this context, broadband near-infrared (NIR) photonics offers a stringent stress test for high-entropy perovskites.[21] Wide-coverage emitters are attractive for spectroscopy, imaging and sensing, yet under realistic operation the output fidelity is often limited not by the nominal bandwidth but by stability.[22] Moisture/oxygen exposure and ion-migration-assisted compositional evolution readily translate into intensity drift and spectral reshaping, compromising quantitative readouts.[23, 24] A platform that can deliver both spectrally orthogonal NIR channels and mechanistically accountable robustness would therefore provide a particularly discriminating benchmark for entropy engineering.[11, 25]

Here we establish entropy-engineered rare-earth halide double perovskite (RHDP) single crystals, $Cs_2Na(Sb, RE)Cl_6$ ($RE^{3+}$ = $Sc^{3+}$, $Er^{3+}$, $Yb^{3+}$ and $Tm^{3+}$), as a model platform that simultaneously delivers expanded photonic functionality and provides mechanistic insight into environmental robustness. Near-equiatomic B(III)-site alloying yields a single-phase high-entropy solid solution (configurational entropy, $\Delta S_{config} \approx 1.6R$) with homogeneous multication incorporation. $Sb^{3+}$ acts as a broadband sensitizer that unifies excitation and cooperatively activates multiple lanthanide emission channels, converting the parent single-mode response into a wide-coverage NIR output spanning ~850–1600 nm with three spectrally separated fingerprint bands at 996, 1220, and 1540 nm. This tri-peak, self-referenced signature enables redundancy-based ratiometric analysis that is intrinsically less sensitive to intensity drift, supporting reliable solvent identification and quantitative mixture sensing. Beyond functional expansion, accelerated aging tests reveal markedly enhanced tolerance to humidity and oxygen relative to single-component analogues. Crucially, simulations disentangle the origin of this robustness: surface adsorption calculations indicate that the high-entropy configuration does not significantly reduce $H_2O/O_2$ affinity, consistent with a "cocktail effect" that preserves adsorption capability, whereas molecular dynamics reveals substantially suppressed self-diffusion in the high-entropy lattice (including reduced halide transport). Together with the $-T\Delta S_{config}$ stabilization term, this sluggish transport kinetically impedes long-range ionic migration and lattice reconstruction, delaying the propagation of surface-initiated reactions and thereby enhancing environmental robustness. Finally, integration into a 340 nm–pumped phosphor-converted Light Emitting Diodes (LEDs) yields spectrally stable, wide-coverage NIR illumination, underscoring configurational-entropy engineering as a practical route to concurrently expand photonic functionality and strengthen operational stability in halide-perovskite photonics.

## Results and Discussion

### Synthesis and Structural Characterizations of Pristine and high-entropy RHDPs

Here, high-quality pure $Cs_2NaRECl_6$, as well as $Cs_2Na(Sb_{0.2}Sc_{0.2}Er_{0.2}Yb_{0.2}Tm_{0.2})Cl_6$ (abbreviated as HEA-Yb-Tm-Er), were synthesized using an adapted hydrothermal method.[26] Typically, powder reagents including cesium chloride, sodium chloride, and rare earth hydrated chlorides were loaded into a 25-mL Teflon vessel, followed by the addition of concentrated hydrochloric acid (See the experimental part for details). Crystallization proceeds under a precisely controlled thermal profile (180 ± 0.5 °C for 12 h) followed by slow gradient cooling (5 °C/h). In the hydrothermal growth of rare earth-based high-entropy halide perovskites (HEHPs), the slow programmed cooling stage (e.g., 5 °C/h) plays an independent and critical role. Its essence lies in the continuous and precise control of the system's supersaturation, keeping it within the kinetic metastability window, thereby suppressing explosive nucleation and driving the oriented growth of a limited number of crystal nuclei. This step is beneficial for high-entropy growth: slow cooling provides time for multication diffusion/equilibration, enabling homogeneous incorporation and B-site ordering in the double-perovskite lattice, while Ostwald ripening heals defects. It also drives the system toward the thermodynamic ground state, suppressing metastable or secondary phases. Consequently, programmed cooling governs crystal size, compositional uniformity, structural integrity, and phase purity.

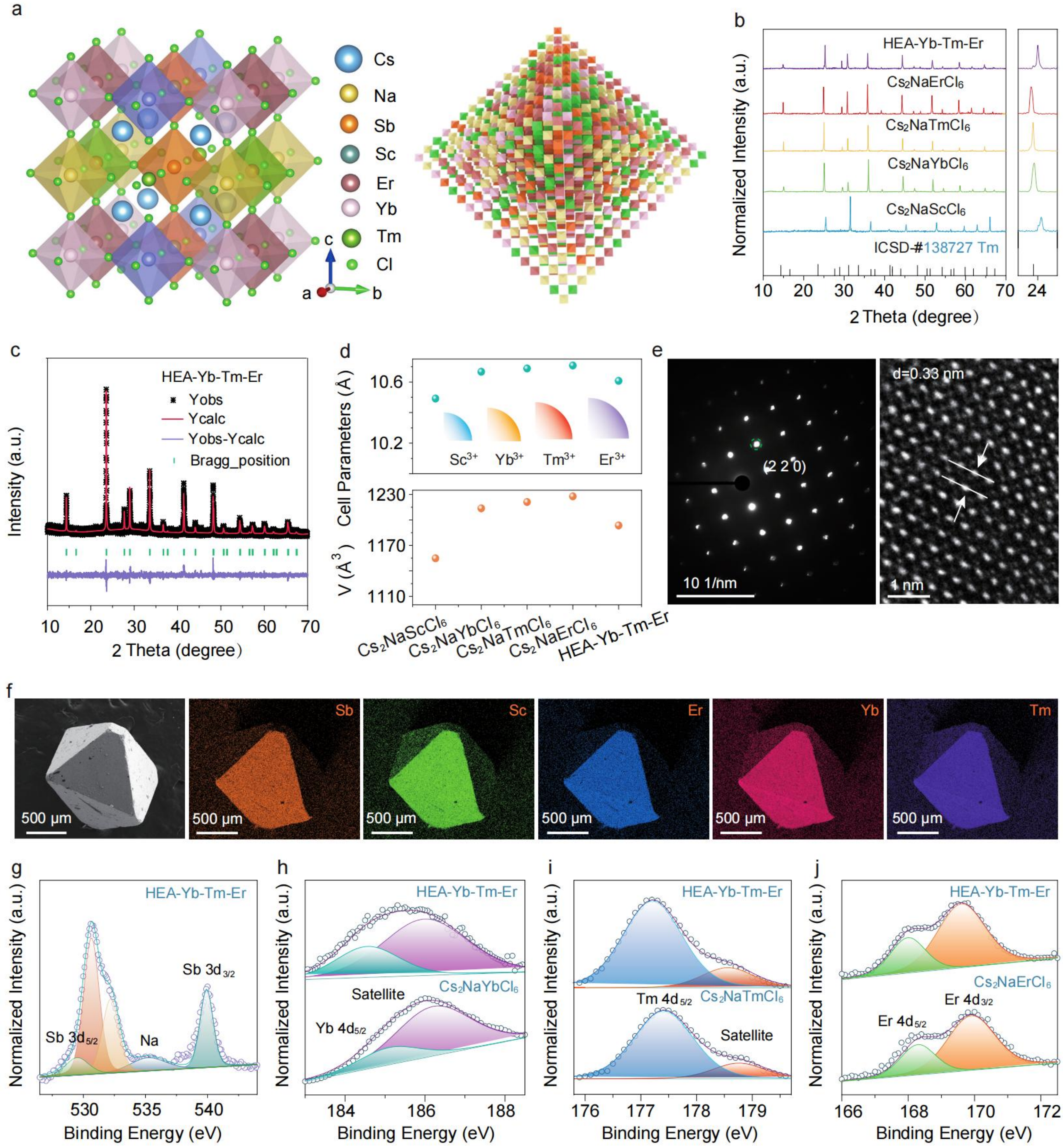

**Fig. 1 | Structural and compositional characterization of HEA-Yb-Tm-Er single crystals. a,** Schematic of high-entropy double perovskite structure with five randomly distributed $Sb^{3+}$ and $RE^{3+}$ cations ($Sc^{3+}$, $Yb^{3+}$, $Tm^{3+}$, $Er^{3+}$). **b,** PXRD patterns of HEA-Yb-Tm-Er and parent $Cs_2NaRECl_6$ ($RE^{3+}$ = $Sc^{3+}$, $Tm^{3+}$, $Er^{3+}$, $Yb^{3+}$) single crystals confirm single-phase structure. **c,** Rietveld refinement of HEA ($\chi^2$ = 1.16, Rwp = 6.3%) based on *Fm-3m* model confirms structural consistency with double perovskite lattice. **d,** Lattice parameters of parent crystals follow a monotonic trend with ionic radius; HEA-Yb-Tm-Er lies near the average, indicating effective multicomponent incorporation. **e,** SAED and HRTEM of HEA-Yb-Tm-Er show clear lattice fringes with 0.33 nm d-spacing, corresponding to (220) planes. **f,** SEM and EDS mapping show uniform spatial distribution of all elements. **g, h, i,**

**j,** High-resolution XPS spectra of, Sb 3d (**g**), Yb 4d (**h**), Tm 4d (**i**) and Er 4d (**j**) orbitals in $Cs_2NaRECl_6$ and HEA-Yb-Tm-Er single crystals.

We design HEHPs starting from the representative $Cs_2NaRECl_6$ by introducing $Sb^{3+}$, $Sc^{3+}$, $Er^{3+}$, $Yb^{3+}$ and $Tm^{3+}$ into the B(III) sites. To verify the success of synthesis, we first examined the phase purity and average crystal structure. The HEA-Yb-Tm-Er single crystals crystallize in a cubic system with the *Fm-3m* space group, exhibiting high symmetry (Fig. 1a). The lattice consists of corner-sharing $[NaCl_6]^{5-}$, $[SbCl_6]^{3-}$ and $[RECl_6]^{3-}$ octahedra, with $Cs^+$ cations occupying the cavities formed by these octahedral units. The powder X-ray diffraction (PXRD) patterns of the as-prepared samples (Fig. 1b) confirm the RHDP structure, as evidenced by the close match with the standard $Cs_2NaTmCl_6$ reference pattern (ICSD No. 138727). Because $Er^{3+}$ ($Yb^{3+}$, $Sc^{3+}$) has a larger (smaller) ionic radius than $Tm^{3+}$, the diffraction peaks of $Cs_2NaErCl_6$ ($Cs_2NaYbCl_6$, $Cs_2NaScCl_6$) shift to lower (higher) angles in the enlarged $2\theta$ range of 23–25°, consistent with lattice expansion and contraction. Upon multication substitution at the B(III) site, HEA-Yb-Tm-Er shows a slight peak shift, which can be attributed to the different ionic radii of $Sb^{3+}$ (0.76 Å), $Sc^{3+}$ (0.74 Å), $Tm^{3+}$ (0.88 Å), $Yb^{3+}$ (0.87 Å) and $Er^{3+}$ (0.89 Å).[27] This behavior indicates a single-phase composition at the macroscopic scale.

To further clarify the structure, Rietveld refinements were carried out on the XRD data of both $Cs_2NaRECl_6$ and HEA-Yb-Tm-Er single crystals (Fig. 1c, Supplementary Fig. S1 and Table S1). All refinements yield reliable goodness-of-fit factors ($\chi^2 < 1.5$, Rp and Rwp < 10%), confirming the robustness of the models. The variations in unit-cell parameters (a, b, c) and volume (V) for $Cs_2NaRECl_6$ and HEA-Yb-Tm-Er correlate well with the corresponding XRD peak shifts (Fig. 1d). These results are consistent with a homogeneous single-phase solid solution in which all cations are randomly distributed over equivalent lattice sites. The single-crystal nature and lattice periodicity of HEA-Yb-Tm-Er were then verified at the microscopic level. The selected-area electron diffraction (SAED) pattern shows sharp diffraction spots (Fig. 1e left), indicating good crystallinity. High-resolution transmission electron microscopy (HRTEM) images reveal clear lattice fringes with a spacing of 0.33 nm, corresponding to the (220) plane. This spacing is smaller than that of $Cs_2NaErCl_6$, confirming lattice contraction induced by $Sb^{3+}$ and $RE^{3+}$ alloying (Fig. 1e right).

We next evaluated the multicomponent stoichiometry and its spatial uniformity. Scanning electron microscopy (SEM) and energy-dispersive spectrometer (EDS) elemental mapping (Supplementary Figs. S2–S5 and Fig. 1f) show that Cs, Na, Sb, Cl, Sc, Yb, Tm and Er are homogeneously distributed throughout individual HEA-Yb-Tm-Er crystals, without detectable phase segregation. Quantitative EDS analysis gives an atomic ratio of Cs/Na/Sb/Sc/Tm/Yb/Er/Cl = 1.98/0.97/0.23/0.18/0.21/0.20/0.19/6.03, which agrees well with the ICP-OES result for HEA-Yb-Tm-Er single crystals (Supplementary Fig. S6 and Table S2). These data confirm that the intended multicomponent composition is realized and that single-phase HEA-Yb-Tm-Er crystals are obtained. For clarity, the feeding ratios are used throughout the discussion. In addition, EDS spectra for $Cs_2NaRECl_6$ ($RE^{3+}$ = $Sc^{3+}$, $Tm^{3+}$, $Er^{3+}$, $Yb^{3+}$) confirm the expected 2:1:1:6 atomic ratio for Cs:Na:RE:Cl (Fig. S7), demonstrating the successful synthesis of $Cs_2NaRECl_6$.

The chemical states of the cations were further probed by X-ray photoelectron spectroscopy (XPS). Comparative XPS measurements (Figs. 1g–1h and Supplementary Fig. S8) show that Sb and all RE ions are present in the +3 oxidation state. The full survey spectrum contains clear signals from Cs 3d, Na 1s, Sb 3d, Sc 2p, Yb 4d, Tm 4d, Er 4d and Cl 2p. As shown in Figs. 1g–1i, the Sb $3d_{3/2}$ and $3d_{5/2}$ peaks appear at 539.9 and 529.5 eV, while the peak at 184.5 eV and 186.1 eV is assigned to the Yb $4d_{5/2}$ level and its satellite peak.[28-30] The peaks at 169.6 and 168.0 eV correspond to the Er $4d_{3/2}$ and $4d_{5/2}$ levels, respectively, those at 177.2 and 178.6 eV are assigned to the Tm $4d_{5/2}$ level along with its satellite.[31-33] Relative to the pure $Cs_2NaRECl_6$ single crystals, these binding energies exhibit systematic shifts, indicating modified local electron densities in the multicomponent lattice, consistent with an electronic-scale "cocktail effect". Finally, the high-entropy nature was quantified by configurational entropy. A randomly mixed solid solution of HEA-Yb-Tm-Er will have a high atomic configurational entropy ($\Delta S_{config}$), given for ideal solutions by[34, 35]

$$\Delta S_{config} = -R \left(\sum_{i=1}^{N} x_i \ln x_i + \sum_{j=1}^{P} y_j \ln y_j\right) \quad (1)$$

where N and P are the numbers of distinct ion types occupying the Na and B(III) sites, respectively, and $x_i$ and $y_j$ are the mole fractions of the $i^{th}$ and $j^{th}$ ions, respectively. High-entropy materials are defined by a configurational entropy greater than 1.5R. Accordingly, HEA-Yb-Tm-Er single crystals ($\Delta S_{config}$ = 1.6R) exhibits a characteristic high-entropy nature.

**Optical Properties of $Cs_2NaRECl_6$ and HEA-Yb–Tm-Er Single Crystals**

To investigate the role of each component in the high-entropy RHDPs in constructing broadband NIR emission, we conducted optical performance characterization. The absorption spectra for these materials are shown in Fig. 2a. Given the focus of this study on NIR luminescence, $Cs_2NaScCl_6$ single crystals (which lack such emission) are not discussed in detail. The photoluminescence excitation (PLE) and Photoluminescence (PL) emission spectra in the visible (Vis) region (Supplementary Fig. S9) are provided, displaying the widely reported blue self-trapped exciton (STE) emission upon 250 nm excitation.[36] The $Cs_2NaRECl_6$ single crystals ($RE^{3+}$ = $Tm^{3+}$, $Er^{3+}$ and $Yb^{3+}$) exhibit distinct absorption features attributed to their respective $RE^{3+}$ ions. In $Cs_2NaTmCl_6$ single crystals, the absorption bands at 471, 691 and 800 nm were assigned to the $^1G_4 \rightarrow ^3H_6$, $^3F_{2,3} \rightarrow ^3H_6$ and $^3H_4 \rightarrow ^3H_6$ transitions of $Tm^{3+}$ ions, respectively.[37] Similarly, $Cs_2NaErCl_6$ single crystals showed absorption bands at 380 nm, 521 nm and 583 nm, corresponding to the $Er^{3+}$ transitions $^4G_{11/2} \rightarrow ^4I_{15/2}$, $^2H_{11/2} \rightarrow ^4I_{15/2}$ and $^4S_{3/2} \rightarrow ^4I_{15/2}$, respectively.[38] The $Cs_2NaYbCl_6$ single crystals exhibit absorption bands at 271, 405 and 980 nm, which is attributed to the transitions of ground state (GS)$\rightarrow Cl^-$-$Yb^{3+}$ charge transfer band (CTB), $^2F_{5/2} \rightarrow$CTB, and $^2F_{7/2} \rightarrow ^2F_{5/2}$, respectively.[39] Furthermore, alongside the characteristic absorption bands of $Tm^{3+}$, $Er^{3+}$ and $Yb^{3+}$ ions, the HEA-Yb-Tm-Er single crystals demonstrate the pronounced absorption bands centered at 268 nm, 282 nm and 300-380 nm, corresponding to the $^1S_0 \rightarrow ^{1,2,3}P_1$ transition of $Sb^{3+}$ ions, respectively.[40, 41]

As shown in Fig. 2b, $Cs_2NaRECl_6$ single crystals demonstrated the single characteristic NIR emissions at 996 nm ($Yb^{3+}$: $^2F_{5/2} \rightarrow ^2F_{7/2}$), 1220 nm ($Tm^{3+}$: $^3H_5 \rightarrow ^3H_6$) and 1540 nm ($Er^{3+}$: $^4I_{13/2} \rightarrow ^4I_{15/2}$), respectively. Intriguingly, HEA-Yb-Tm-Er single crystals exhibit broad blue self-trapped exciton (STE) emission centered at 460 nm and broadband NIR emission spanning from 850 to 1600 nm, integrating the transitions of $Sb^{3+}$, $Tm^{3+}$, $Er^{3+}$ and $Yb^{3+}$ ions.[40] The PLE spectra elucidate the detailed origins of this emission profile. When the emission wavelengths are centered at 996, 1220, 1540 nm and $^3P_1$-$^1S_0$ emission of $Sb^{3+}$, all the PLE spectrum reveals the distinct excitation bands at 340 nm (Supplementary Fig. S10). These results indicate that the NIR emissions derived from energy transfer from the $Sb^{3+}$ to $RE^{3+}$ ions. The normalized PLE spectra of $Cs_2NaRECl_6$ and HEA-Yb-Tm-Er single crystals, monitored at their respective NIR emissions, are shown in Fig. 2c. The spectra revealed a distinct shift in dominant excitation peaks from $RE^{3+}$-associated bands to $Sb^{3+}$-related features. This spectral evolution indicates a modified excitation pathway, in which the incorporated $Sb^{3+}$ ions act as critical sensitizers.[42] By providing a broad absorption band around 340 nm, they enable efficient energy transfer to $RE^{3+}$ ions upon single-wavelength excitation, thereby maximizing the overall luminescence output. Despite the prevalence of nonradiative cross-relaxation in multi-dopant systems, the incorporation of multiple $RE^{3+}$ ions here show no apparent luminescence quenching under the present compositions and excitation conditions. (Supplementary Fig. S11). This is attributed to the key advantage provided by the large unit cell of $Cs_2NaRECl_6$: it increases the average inter-RE distance (to ~7.6 Å), substantially larger than that in representative rare-earth hosts such as $NaYF_4$ (~3.5 Å)—the $1/d^6$ dependence of energy transfer reduces long-range migration to quenching centers.[43, 44] As a result, interionic coupling and long-range excitation exchange among $RE^{3+}$ centers are less pronounced, which helps suppress migration-assisted quenching and preserves multiple emission channels with reduced mutual interference.

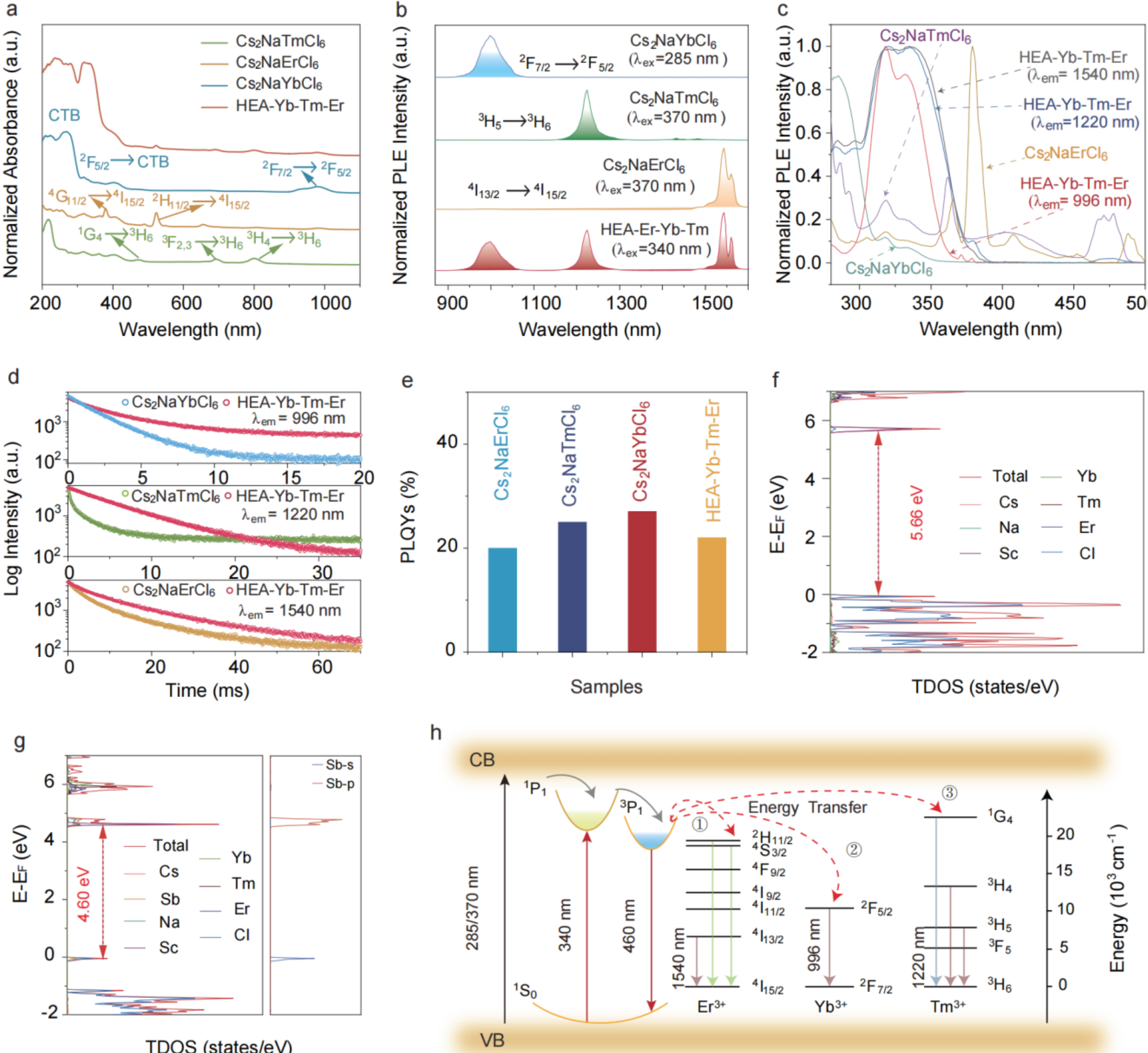

**Fig. 2 | Optical properties of $Cs_2NaRECl_6$ and HEA-Yb–Tm–Er single crystals. a,** UV–Vis–NIR absorption spectra of $Cs_2NaRECl_6$ ($RE^{3+}$= $Er^{3+}$, $Yb^{3+}$, $Tm^{3+}$) and HEA-Yb-Tm-Er single crystals. **b,** NIR PL spectra of $Cs_2NaYbCl_6$, $Cs_2NaTmCl_6$, $Cs_2NaErCl_6$, and HEA-Yb-Tm-Er. **c,** Normalized PLE spectra monitored at different emission wavelengths (996, 1220, and 1540 nm), illustrating excitation pathways for individual $RE^{3+}$ centers in both

parent and HEA crystals. **d,** Time-resolved PL decay curves of $Cs_2NaRECl_6$ and HEA-Yb-Tm-Er single crystals, recorded at emission wavelengths of 996 nm ($Yb^{3+}$), 1220 nm ($Tm^{3+}$), and 1540 nm ($Er^{3+}$). **e,** Absolute PLQYs of $Cs_2NaRECl_6$ and HEA-Yb-Tm-Er single crystals under UV excitation. **f,** PDOS for $Cs_2Na(Sc_{0.25}Yb_{0.25}Tm_{0.25}Er_{0.25})Cl_6$. **g,** PDOS for HEA-Yb-Tm-Er. **h,** Schematic illustration of the PL processes and energy-transfer pathways among $Er^{3+}$, $Yb^{3+}$, and $Tm^{3+}$ ions in HEA-Yb-Tm-Er single crystals.

Time-resolved PL decays were collected for $Cs_2NaRECl_6$ and HEA-Yb-Tm-Er single crystals to probe the deactivation dynamics of the $RE^{3+}$-centered NIR emissions. All samples show biexponential decays that can be described by a fast and a slow component ($\tau_1$ and $\tau_2$), reflecting the coexistence of multiple relaxation channels. Compared with $Cs_2NaRECl_6$ single crystals, the HEA-Yb-Tm-Er single crystals exhibited the attenuated nonradiative and the promoted radiative recombination, as evidenced by their PL lifetimes (Fig. 2d, Supplementary Table S3). Compared with the parent $Cs_2NaRECl_6$ crystals, HEA-Yb-Tm-Er exhibits prolonged effective lifetimes for the $RE^{3+}$-centered NIR emissions, consistent with a reduced rate of long-range energy/ion transport in the compositionally complex lattice, as reported for other high-entropy halide perovskites.[45] The measured Photoluminescence Quantum Yield (PLQY) are 19.9%, 24.9% and 25.1% for $Cs_2NaErCl_6$, $Cs_2NaTmCl_6$ and $Cs_2NaYbCl_6$ single crystals, respectively (Fig. 2e). Although $Sb^{3+}$ provides a broadband absorption channel and enables sensitization of $RE^{3+}$ NIR emission, the PLQY of HEA-Yb-Tm-Er remains ~22.0%. Notably, the longer effective lifetimes without a concomitant PLQY increase suggest that the slowed transport is accompanied by additional parallel deactivation pathways in the entropy-distorted lattice and/or excitation redistribution among multiple competing pathways, which together constrain the net NIR radiative yield.

To further verify the spectral tunability of NIR emissions in the high-entropy system, we synthesized single-luminescent-center $Cs_2Na(Sb_{0.2}Sc_{0.2}Y_{0.2}Gd_{0.2}Tm_{0.2})Cl_6$ (HEA-Tm), $Cs_2Na(Sb_{0.2}Sc_{0.2}Y_{0.2}Gd_{0.2}Er_{0.2})Cl_6$ (HEA-Er), and $Cs_2Na(Sb_{0.2}Sc_{0.2}Y_{0.2}Gd_{0.2}Yb_{0.2})Cl_6$ (HEA-Yb) single crystals and investigated their NIR emission characteristics. The structural characterizations and optical properties are presented in Supplementary Figs. S12-S13. Under 340 nm excitation, HEA-Tm, HEA-Er, and HEA-Yb single crystals exhibited characteristic NIR emissions identical to those of $Cs_2NaRECl_6$ (Supplementary Figs. S14-S16). In contrast, the strategy of multi-emissive-center high-entropy systems represents a paradigm shift. Its value stems from the fundamental combination of entropy-stabilized phases and cooperative elemental effects, which in our system collectively unlock superior and readily tunable broadband emission alongside designer functionalities. The NIR emission lifetimes in HEA-Tm, HEA-Er, and HEA-Yb single crystals followed the same trend as those in HEA-Yb-Tm-Er single crystals (Supplementary Fig. S17), with detailed fitting parameters summarized in Table S3.

Furthermore, we conducted first-principles calculations on $Cs_2NaSbCl_6$, $Cs_2NaRECl_6$ ($RE^{3+}$ = $Sc^{3+}$, $Tm^{3+}$, $Er^{3+}$, $Yb^{3+}$) and HEA-Yb-Tm-Er single crystals to investigate the potential impact of $Sb^{3+}$ and $RE^{3+}$ ions alloying in multicomponent HEA-Yb-Tm-Er single crystals on their electronic band structures and optical properties.[42] We first investigated the electronic structure using multiple computational methods to estimate the band gap. Although different exchange–correlation functionals can yield different absolute band-gap values, semilocal GGA-level approaches such as PBE are widely used as a consistent baseline for comparing trends and for ensuring comparability with prior reports. To balance literature compatibility with improved quantitative accuracy, we adopted a tiered workflow. We first performed PBE calculations to benchmark our structural models and to provide a reference set consistent with earlier studies (Supplementary Fig. S18). We then employed the hybrid functional Heyd–Scuseria–Ernzerhof (HSE06) to refine the electronic structure and obtain band gaps at a higher level of theory.[46-49]

Parent $Cs_2NaSbCl_6$ and $Cs_2NaRECl_6$ ($RE^{3+}$ = $Sc^{3+}$, $Tm^{3+}$, $Er^{3+}$, $Yb^{3+}$) display the band gap of 3.89, 5.56, 6.77, 6.72 and 6.76 eV, respectively (Supplementary Fig. S19 left). The projected density of states (PDOS) analysis indicates that the valence band maximum (VBM) is dominated by Cl 3p orbitals, while the conduction band minimum (CBM) comprises a mixture of Sb 5p, Sc 3d, Tm 4d, Er 4d, and Yb 4d orbitals, along with Cl 3p states (Supplementary Fig. S19 right). To elucidate the sensitization role of Sb in the multicomponent HEA-Yb-Tm-Er, we also computed the PDOS of the $Cs_2Na(Sc_{0.25}Yb_{0.25}Tm_{0.25}Er_{0.25})Cl_6$. Comparative PDOS analysis reveals that $Sb^{3+}$ doping creates a new sensitization center by introducing Sb 5s states above the VBM and Sb 5p states below the CBM. The feasible 5s–5p optical transitions of this center enable efficient energy transfer within the HEA-Yb-Tm-Er single crystals. Guided by above experimental and first principles calculations results, the mechanism is proposed that incorporation of $Sb^{3+}$ ions unified the absorption channel, establishing an efficient photoexcitation-recombination in HEA-Yb-Tm-Er single crystals (Fig. 2h). Upon excitation at 340 nm, the $Sb^{3+}$ ion was excited from the $^1S_0$ to the $^3P_1$ state, which then underwent dynamic Jahn–Teller distortion, resulting in broad blue emission centered at 460 nm via the $^3P_1 \rightarrow ^1S_0$ transition. Simultaneously, part of the STE energy was non-radiatively transferred to nearby $Yb^{3+}$, $Tm^{3+}$ and $Er^{3+}$ ions, giving rise to intense NIR emissions at 996, 1220 and 1540 nm. This process collectively achieves a broadband NIR emission spanning approximately 850-1600 nm.

**Environmental Stability Assessment of $Cs_2NaRECl_6$ and HEA-Er–Yb–Tm single crystals**

Given the well-documented environmental sensitivity of perovskite materials, we systematically evaluated $Cs_2NaRECl_6$ ($RE^{3+}$ = $Sc^{3+}$, $Tm^{3+}$, $Er^{3+}$, $Yb^{3+}$) and HEA-Yb-Tm-Er single crystals under harsh conditions of prolonged high-humidity and oxygen exposure, with the dual aims of assessing their robustness and elucidating the role of the high-entropy effect. As moisture represents a critical environmental stressor, we first examined the behavior under controlled humidity. All samples were subjected to accelerated aging tests under controlled high-humidity conditions (100% RH) at ambient temperature (25-30°C), with periodic XRD characterization to analyze the structural evolution. After being subjected to a 12-hour exposure, $Cs_2NaTmCl_6$ exhibited additional diffraction peaks corresponding to NaCl and $TmCl_3$ in its XRD pattern (Fig. 3a), suggesting the onset of partial decomposition. With prolonged exposure time, the XRD peaks of NaCl and $TmCl_3$ intensified while the crystalline peaks of $Cs_2NaTmCl_6$ gradually weakened. After 24 hours under high relative humidity (100% RH), the diffraction intensities of impurity phases became comparable to those of $Cs_2NaTmCl_6$, indicating near-complete sample decomposition. Similar degradation behavior was observed for other $Cs_2NaRECl_6$ ($RE^{3+}$ = $Sc^{3+}$, $Er^{3+}$, and $Yb^{3+}$) single crystals under identical conditions, with extensive decomposition occurring within 24 hours (Supplementary Fig. S20).

In contrast, the HEA-Yb-Tm-Er single crystals demonstrated enhanced phase stability under high-humidity conditions, as their XRD patterns showed only minimal NaCl impurity formation after 24-hour exposure under identical conditions (Fig. 3b). Furthermore, the moisture adsorption characteristics of both $Cs_2NaRECl_6$ ($RE^{3+}$ = $Sc^{3+}$, $Er^{3+}$, and $Yb^{3+}$) and HEA-Yb-Tm-Er single crystals were compared using gravimetric analysis (Fig. 3c). The HEA crystals exhibit a substantially smaller mass increase over time, indicating a reduced overall moisture uptake under identical high-humidity exposure.

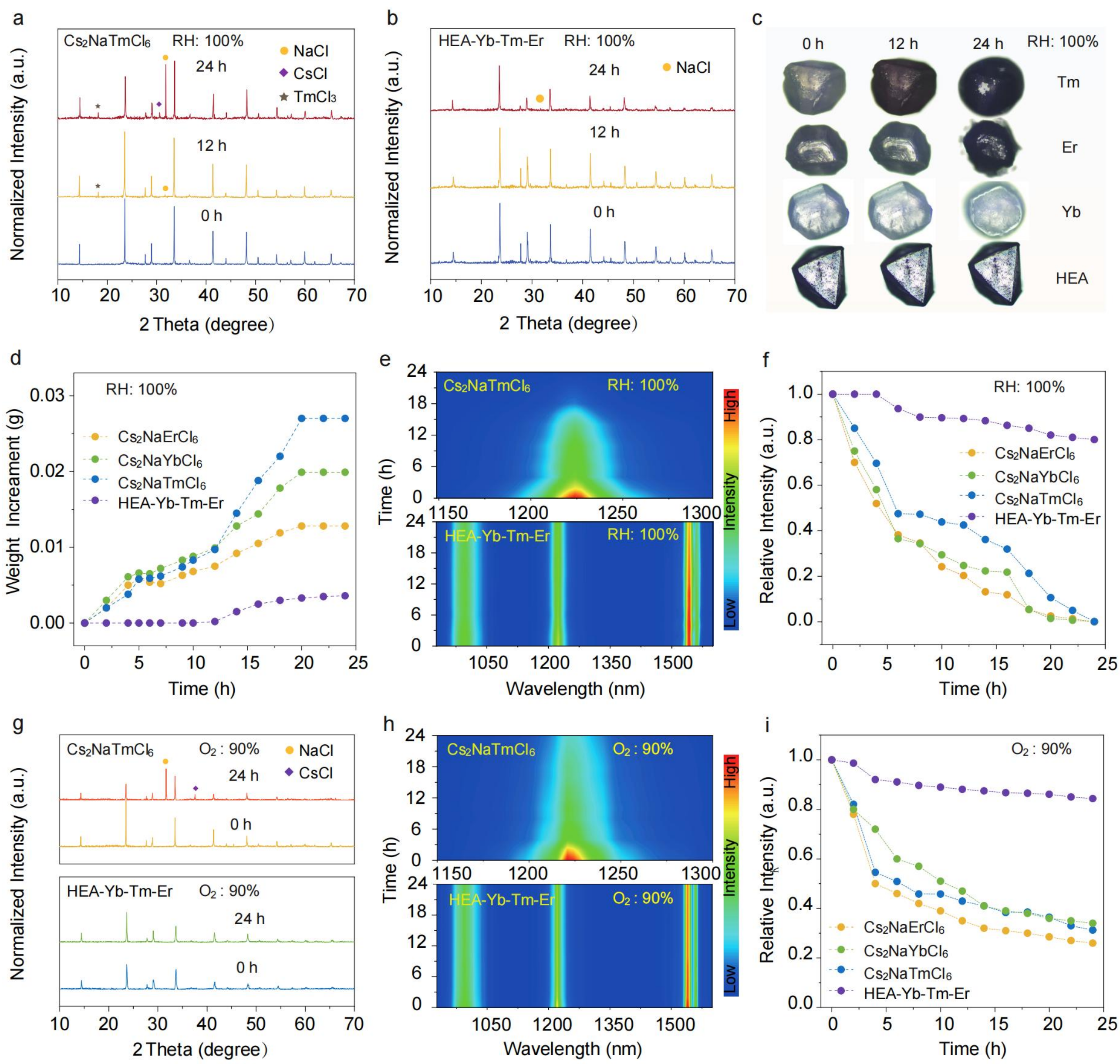

**Fig. 3 | Structural and optical stability evaluation of $Cs_2NaRECl_6$ ($RE^{3+}$ = $Tm^{3+}$, $Er^{3+}$, $Yb^{3+}$) and HEA-Yb-Tm-Er single crystals under humid and oxidative conditions. a, b,** XRD patterns of $Cs_2NaTmCl_6$ (**a**) and HEA-Yb-Tm-Er (**b**) single crystals after 0, 12, and 24 h exposure at 100% RH. **c,** Optical images of the crystals under 100% RH at different durations. **d,** Time-dependent $H_2O$ absorption measured by weight gain. **e,** Evolution of NIR PL spectra of $Cs_2NaTmCl_6$ (top) and HEA-Yb-Tm-Er (bottom) under 100% RH. **f,** Relative NIR PL intensity as a function of time for all samples under 100% RH. XRD. **g,** and time-resolved NIR PL spectra. **h,** of $Cs_2NaTmCl_6$ (top) and HEA-Er-Yb-Tm (bottom) under oxygen atmosphere ($O_2$, 90%; 50–60% RH). **i,** Time-dependent PL degradation under $O_2$ for all samples. All XRD measurements for humidity stability tests were performed after a drying pretreatment to eliminate adsorbed $H_2O$ molecules.

The optical microscopy images in Fig. 3d and Supplementary Fig. S21 document the hygroscopic degradation process of $Cs_2NaRECl_6$ ($RE^{3+} = Sc^{3+}$, $Tm^{3+}$, $Er^{3+}$, $Yb^{3+}$) and HEA-Yb-Tm-Er single crystals under high-humidity conditions over 0-24 hours. The degradation of the $Cs_2NaRECl_6$ under high humidity follows a well-defined sequence, progressing from initial surface attack to complete structural collapse. Ultimately, cumulative chemical and mechanical damage leads to loss of macroscopic integrity (as evidenced by the impurity phase XRD peaks), transforming the well-defined single crystal into a fragmented state. In striking contrast, the HEA-Yb-Tm-Er single crystals maintained structural integrity throughout the 24-hour test, showing neither large amount of $H_2O$ adsorption nor surface morphological changes. These results indicated that the high configurational entropy inherent to the high-entropy design thermodynamically stabilizes the single-phase solid-solution. In a humid environment, this stabilization effect enables the HEA-Yb-Tm-Er single crystals to resist the nucleation and growth of degradation-prone, moisture-sensitive impurity phases.

To further investigate whether the enhanced humidity stability is entropy-driven, we conducted comparative humidity stability tests (100% RH, 25-30°C) on HEA-Tm, HEA-Er, and HEA-Yb single crystals, respectively. The XRD characterization demonstrated that they exhibited similar stability, showing weak NaCl impurity peaks after 24-hour exposure (Supplementary Fig. S22). The HEA-Tm, HEA-Er, and HEA-Yb single crystals exhibited significantly enhanced moisture resistance compared to $Cs_2NaRECl_6$ ($RE^{3+} = Tm^{3+}$, $Er^{3+}$, $Yb^{3+}$) counterparts (Supplementary Fig. S23). These collective results confirm configurational entropy stabilization as the dominant mechanism underlying the exceptional environmental stability of high-entropy alloy single crystals.

The NIR PL spectra of $Cs_2NaRECl_6$ ($RE^{3+} = Tm^{3+}$, $Er^{3+}$, $Yb^{3+}$), HEA-RE ($RE^{3+} = Tm^{3+}$, $Er^{3+}$, $Yb^{3+}$), and HEA-Yb-Tm-Er single crystals were measured after 24-hour exposure to controlled humidity conditions (100% RH, 25°C). It can be observed that the NIR PL intensity of $Cs_2NaTmCl_6$ single crystals exhibited a progressive decrease during humidity exposure, remaining only ~2 % of its initial PL intensity after 24 hours (Fig. 3e top and Fig. 3f). Similarly, the NIR PL emissions of both $Cs_2NaYbCl_6$ and $Cs_2NaErCl_6$ single crystals showed complete quenching after 24-hour exposure under identical high-humidity conditions (100% RH, 25°C) (Supplementary Fig. S24 and Fig. 3f). The structural degradation directly correlates with a significant deterioration in optical performance. Notably, HEA-Er-Yb-Tm and HEA-RE ($RE^{3+} = Tm^{3+}$, $Er^{3+}$, $Yb^{3+}$) single crystals retained 80%, 40%, 21%, and 17% of their initial NIR PL intensities after 24-hour humidity exposure, respectively (Figs. 3e bottom, 3f, and Supplementary Fig. S25). Combined structural characterization and PL studies confirm that entropy stabilization effects significantly improve the humidity resistance of RHDPs materials.

The potential application of HEA-Yb-Tm-Er single crystals in ambient NIR photonics necessitates a fundamental understanding of its degradation pathways in air. Beyond single-factor tests, the interplay between oxygen and moisture is often decisive for long-term performance. Therefore, we investigate the synergistic effects of elevated $O_2$ concentration (90%) and approximately 50-60% RH on the HEA-Yb-Tm-Er single crystals stability. XRD analysis (Fig. 3g, Supplementary Figs. S26-S27) reveals impurity phase diffraction peaks in $Cs_2NaRECl_6$ ($RE^{3+}$ = $Tm^{3+}$, $Er^{3+}$, $Yb^{3+}$) single crystals, while HEA-RE ($RE^{3+}$ = $Tm^{3+}$, $Er^{3+}$, $Yb^{3+}$) and HEA-Yb-Tm-Er systems remain phase-pure. After 24-hour exposure, the NIR PL intensities of $Cs_2NaRECl_6$ ($RE^{3+}$ = $Tm^{3+}$, $Er^{3+}$, $Yb^{3+}$), HEA-RE ($Re^{3+}$ = $Tm^{3+}$, $Er^{3+}$, $Yb^{3+}$), and HEA-Yb-Tm-Er single crystals retained 31%, 26%, 34%, 71%, 63%, 58%, and 84% of their initial values, respectively (Figs. 3h-3i, Supplementary Figs. S28-S29). It is further confirmed by the measured results that high configurational entropy stabilizes HEA-Yb-Tm-Er single crystals against decomposition and oxygen-induced reactions. However, the mechanistic origin of the enhanced environmental stability of HEA–Er–Yb–Tm is not yet fully resolved and is examined in detail in the theoretical analysis section.

**Theoretical insights into the underlying mechanisms of enhanced stability of HEA-Yb-Tm-Er single crystals**

The enhanced stability of the HEA-Yb-Tm-Er single crystals was evaluated by calculating their decomposition enthalpy ($\Delta H_{decomposition}$) and decomposition Gibbs free energy ($\Delta G_{decomposition}$) as a measure of thermodynamic stability based on DFT. Fig. 4a presents the ΔH of six random configurations of HEA-Yb-Tm-Er single crystals (320 atoms), with values ranging up to 139.2 meV/f.u.. The $T\Delta S_{config}$ for HEA-Yb-Tm-Er single crystals are –35.6 meV. Consequently, the ΔG, derived from $\Delta G = \Delta H - T\Delta S_{config}$, reaches a maximum value of 174.8 meV/f.u.. For the pure $Cs_2NaSbCl_6$ and $Cs_2NaRECl_6$ ($RE^{3+}$ = $Sc^{3+}$, $Tm^{3+}$, $Yb^{3+}$, and $Er^{3+}$), the ΔH are 108.1, 128.9, 141.5, 154.0, and 174.0 meV/f.u., respectively. Due to their $T\Delta S_{config}$ is 0 meV, the corresponding ΔG are identical to the ΔH values. A comparison reveals that HEA-Yb-Tm-Er single crystals possess a larger (more positive) ΔG than pure $Cs_2NaSbCl_6$ and $Cs_2NaRECl_6$ single crystals. Consequently, the more positive ΔG of HEA-Yb-Tm-Er single crystals indicates that its decomposition is less spontaneous, which correlates with enhanced stability. DFT calculations were performed to calculate the adsorption energies of $H_2O$ and $O_2$ on Na and different RE sites (RE = Sc, Tm, Yb, and Er) of the (001) facet of HEA-Yb-Tm-Er, $Cs_2NaSbCl_6$, and $Cs_2NaRECl_6$ ($RE^{3+}$ = $Sc^{3+}$, $Tm^{3+}$, $Yb^{3+}$, and $Er^{3+}$) (Fig.4b and Supplementary Figs. S30-S31, Table S4). The calculated results indicate that both $H_2O$ and $O_2$ exhibit a specific affinity to the surface of HEA-Yb-Tm-Er. Unexpectedly, the high-entropy configuration does not significantly diminish the affinity of the surface, but rather, it manifests the 'cocktail effect', maintaining the adsorption capacity similar to that of the pure phase.

Although surface adsorption energy is a critical parameter for evaluating stability, static calculations based on it are insufficient to capture the dynamic intricacies degradation processes. To address this, we performed a series of molecular dynamics (MD) simulations to investigate the real-time structural evolution. Specifically, to probe the environmental stability of HEA-Yb-Tm-Er and $Cs_2NaRECl_6$ ($RE^{3+}$ = $Tm^{3+}$, $Yb^{3+}$, and $Er^{3+}$), a series of simulations were conducted by introducing a $H_2O$-$O_2$-rich environment to the upper and lower surfaces of the supercell. A simulation temperature of 500 K was selected to accelerate the degradation kinetics of room temperature (300 K). A part of representative results is shown in the main text, others are displayed in the supporting information. Fig. 4c illustrates the structure evolution of the HEA-Yb-Tm-Er (top) and the pure-phase $Cs_2NaTmCl_6$ (bottom) at different MD simulation duration of 0, 5, 20, 30, and 50 ps. For the pure-phase $Cs_2NaTmCl_6$, severe lattice distortion and the collapse of surface octahedra occurred rapidly between 5 and 20 ps. An analogous structural collapse phenomenon, involving severe lattice distortion and the disintegration of surface octahedra, was also observed in both $Cs_2NaYbCl_6$ and $Cs_2NaErCl_6$ within a similar timeframe (0–50 ps) (Supplementary Fig. S32). In contrast, the skeleton of HEA-Yb-Tm-Er maintained quite well with significantly longer duration in the simulation, the severe lattice distortion appeared only after 50 ps annealing. This demonstrates the superior stability of the high-entropy structure under exposure to moisture and oxygen. The lattice distortion of HEA-Yb-Tm-Er and $Cs_2NaTmCl_6$ are clearly quantified via the evolution of Cl-Tm-Cl angle in the $[TmCl_6]^{3-}$ octahedra during the MD simulation in Fig. 4d, from where one can clearly see that the distortion of pure phase $Cs_2NaTmCl_6$ (ca. 30°) is about three times larger than that of HEA-Yb-Tm-Er (ca. 10°). Compared to the HEA-Yb-Tm-Er, the pure-phase $Cs_2NaTmCl_6$ structure exhibited a rapid decrease of Cl-Tm-Cl angle (deviation from 180°), which directly relate to the collapse of the framework observed. Similar degradation trends were observed for the pure phase $Cs_2NaErCl_6$ and $Cs_2NaYbCl_6$ (Supplementary Fig. S33), reinforcing the conclusion that the HEA-Yb-Tm-Er configuration offers significantly enhanced stability compared to its single-component counterparts.

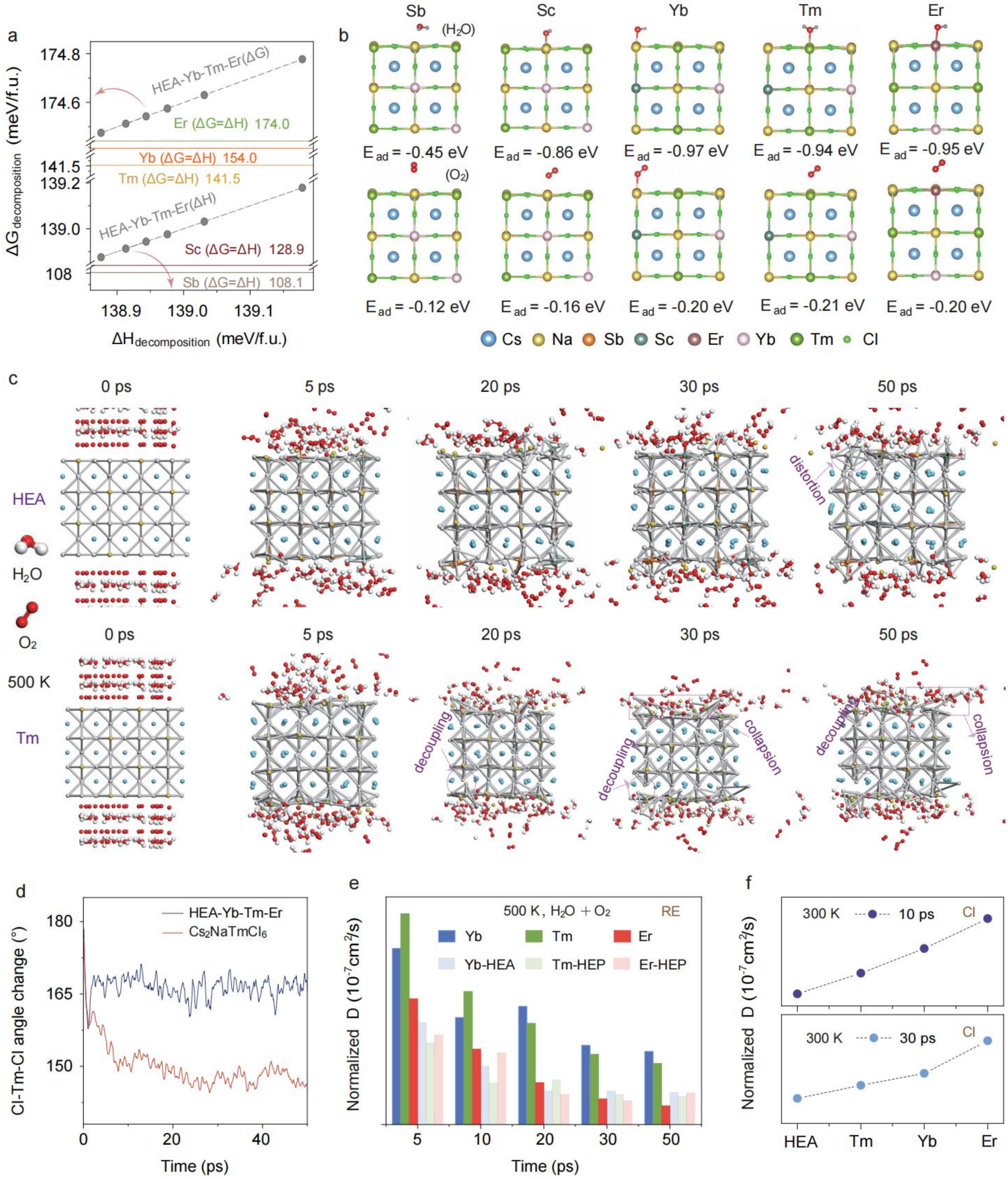

**Fig. 4 | DFT insights into the mechanisms of enhanced stability of HEA-Yb-Tm-Er single crystals. a,** Comparative profiles of $\Delta H_{decomposition}$ and $\Delta G_{decomposition}$ for the six systems studied. Data are presented as: gray line (HEA-Yb-Tm-Er), light brown line ($Cs_2NaSbCl_6$), reddish brown line ($Cs_2NaScCl_6$), orange line ($Cs_2NaYbCl_6$), yellow line ($Cs_2NaTmCl_6$), and green line ($Cs_2NaErCl_6$). HEA-Yb-Tm-Er exhibit a more positive $\Delta G_{decomposition}$ than their pure counterparts ($Cs_2NaSbCl_6$ and $Cs_2NaRECl_6$), signifying suppressed decomposition spontaneity and enhanced stability. **b,** Adsorption configurations of $H_2O$ (top) and $O_2$ (bottom) at the Na, Sb, and RE sites of HEA-

Yb-Tm-Er, respectively. **c,** Snapshots of the final frame of HEA-Yb-Tm-Er (top) and $Cs_2NaTmCl_6$ (bottom) molecular dynamic simulation from 0 ps to 50 ps at 500 K in a $H_2O+O_2$ environment. **d,** Comparative dynamics of the Tm–Cl–Tm bond angle over time at 500 K in $H_2O+O_2$, contrasting the parent $Cs_2NaTmCl_6$ with the HEA-Yb-Tm-Er. **e,** Comparison of $RE^{3+}$ (Yb, Tm, Er) self-diffusion at 500 K in $H_2O+O_2$: parent $Cs_2NaRECl_6$ versus HEA-Yb-Tm-Er. The high-entropy phase exhibits reduced mobility for all cations. **f,** Suppressed $Cl^-$ diffusion in HEA-Yb-Tm-Er. The self-diffusion coefficient of $Cl^-$ at 300 K is lower in the HEA-Yb-Tm-Er than in the parent $Cs_2NaRECl_6$ (Yb, Tm and Er), demonstrating the impact of high-entropy effect on ion transport dynamics.

Considering that $H_2O$ and $O_2$ adsorption primarily occurs at the RE sites, the diffusion of $RE^{3+}$ ions behavior becomes a critical indicator of structural integrity. As illustrated in Fig. 4e and Supplementary Figs. S34-S37, the $RE^{3+}$ ions within the high-entropy structure exhibit significantly suppressed diffusion coefficients (D) compared to their pure-phase $Cs_2NaRECl_6$ ($RE^{3+} = Tm^{3+}$, $Yb^{3+}$, and $Er^{3+}$). This phenomenon, attributed to the sluggish diffusion effect inherent in HEA-Yb-Tm-Er, indicates that ions migration is kinetically hindered. Consequently, the restricted $RE^{3+}$ ions mobility effectively prevents structural reconstruction and phase segregation under reaction conditions, thereby substantiating the superior environmental stability of HEA-Yb-Tm-Er. In parallel, we investigated the diffusion dynamics of corrosive $Cl^-$ ions in HEA-Yb-Tm-Er and $Cs_2NaRECl_6$ ($RE^{3+} = Tm^{3+}$, $Yb^{3+}$, and $Er^{3+}$) (Supplementary Fig. S38). Aligning with the behavior of the $RE^{3+}$ ions, the $Cl^-$ diffusion rate in the HEA-Yb-Tm-Er is significantly lower than in the pure-phase structures. This dual suppression of both host lattice ions and extrinsic corrosive species strongly supports the enhanced environmental stability of the HEA-Yb-Tm-Er system.

To further probe the kinetic stability under realistic operating conditions, we calculated the normalized D of $Cl^-$ ions at 300 K (Fig. 4f). The simulation results reveal a distinct trend where the diffusivity of $Cl^-$ ions follow the order: HEA < $Cs_2NaTmCl_6$< $Cs_2NaYbCl_6$ < $Cs_2NaErCl_6$. Notably, this suppression of ion migration in the high-entropy lattice is time-independent, maintaining the lowest D values at both 10 ps and 30 ps intervals compared to the pure-phase counterparts. Such consistently reduced diffusivity is consistent with a more rugged migration landscape in the high-entropy lattice, where local structural distortions can increase the heterogeneity of the ionic migration pathways and thereby hinder long-range halide transport.[50, 51] This result provides robust evidence that the high-entropy strategy significantly enhances the material's resistance to halide-induced degradation.

Together with the $-T\Delta S_{config}$ term, simulations identify transport retardation as a key contributor to the enhanced robustness of HEA-Yb-Tm-Er. The adsorption calculations show comparable $H_2O$-$O_2$ affinity to the parent phases, whereas MD reveals markedly reduced diffusion for both $RE^{3+}$ cations and $Cl^-$ anions in the high-entropy lattice. These results connect entropy engineering to a measurable descriptor—suppressed ionic mobility—providing a mechanistic handle for improving environmental stability.

**Application of broadband NIR Emission**

HEA-Yb-Tm-Er single crystals leverage the distinct optical characteristics of multiple elements, achieving an broadband single-component NIR emission (850–1600 nm). This property prompted us to construct NIR LEDs by combining these single-component, wide-coverage NIR-emitting materials with a 340 nm UV chip. Supplementary Fig. S39 displays the bias-dependent electroluminescence (EL) spectra of the fabricated NIR LED, which demonstrates spectral tunability over an broadband range showing distinct peaks at 996, 1220, and 1540 nm. As the drive voltage increases, only the intensity of the spectra changes, while the spectral profile remains constant, demonstrating excellent spectral stability for practical applications.

The NIR light sources have emerged as a cornerstone of modern photonic technology, owing to their indispensable role in diverse fields ranging from night-vision surveillance and non-destructive food analysis to deep-tissue bioimaging.[52, 53] Fig. 5a schematically illustrates the principle of NIR imaging using an LED, showing how it enables the capture of detailed images. Fig. 5b–5d display a comparison under natural light (top) and NIR-LED illumination (bottom), with corresponding high-contrast, black-and-white NIR camera images shown below, clearly revealing fine details through NIR imaging. When the as-prepared NIR-LED is used for imaging human tissue, the transmitted light reveals the venous blood vessels in the palm, as captured by the NIR camera (Fig. 5c). This contrast is attributed to the differential absorption of specific NIR wavelengths by chromophores in the blood.[54] The capability of NIR spectroscopy for early and non-destructive detection of food spoilage underpins its practical utility. While the apples appear fresh under visible light (Fig. 5d top), the NIR image under pc-LED illumination (Fig. 5d bottom) shows a dark region on one, corresponding to elevated moisture content—an early spoilage indicator. This contrast is attributed to the absorption of NIR light by water hydroxyl groups, which attenuates the reflected signal and implies a reduced shelf life for that apple. The operational stability of the NIR LED was evaluated over an extended period. As shown in Supplementary Fig. S40, the device maintained stable emission intensity with minimal variation during 50 hours of continuous operation. Combined with its intrinsic NIR spectral features—including invisibility to the human eye and good tissue

penetration—this robust performance supports its potential use in night-vision and bioimaging applications.

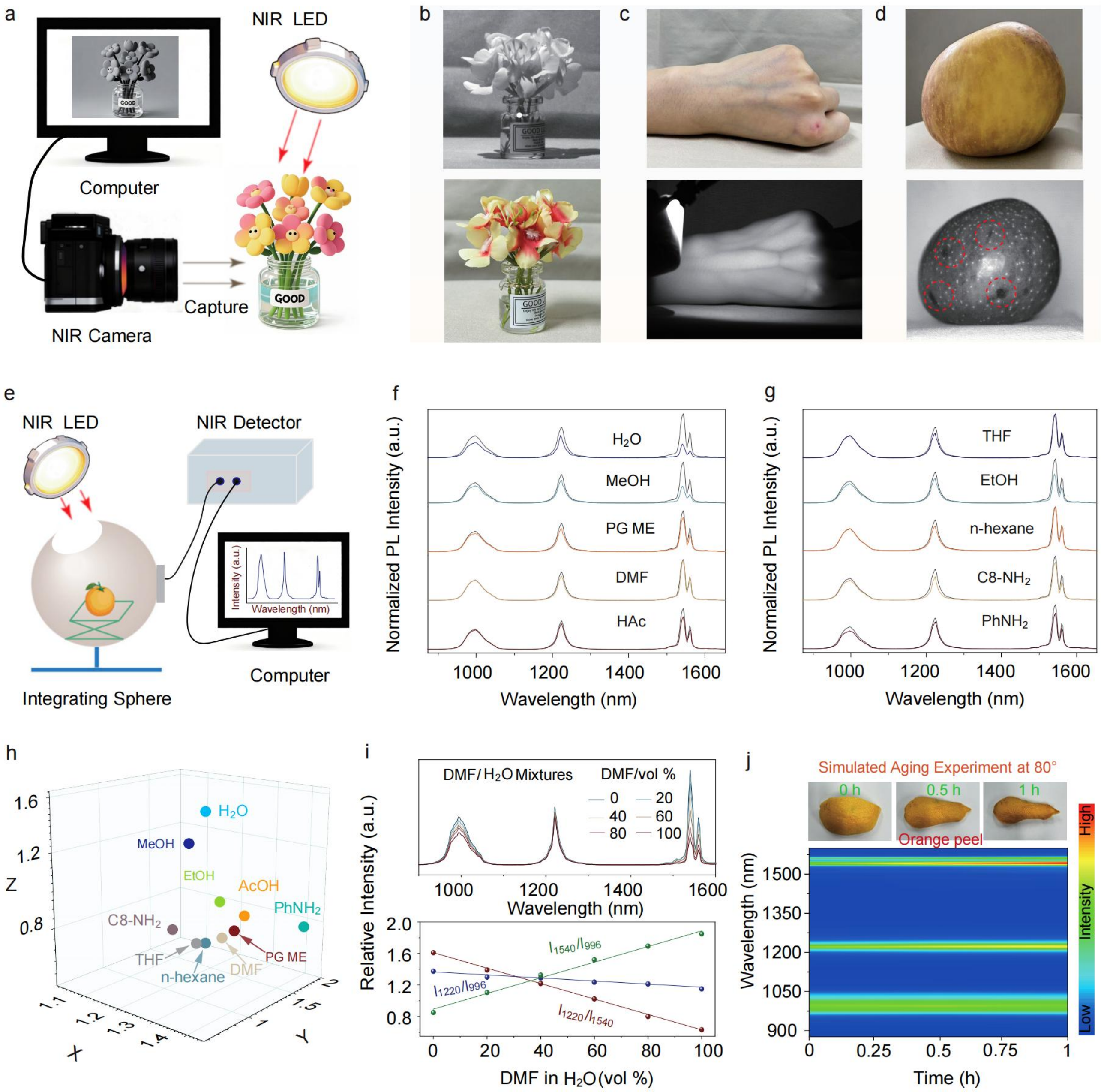

**Fig. 5 | Tri-peak, self-referenced NIR-II photonics enabled by the HEA-Yb-Tm-Er phosphor-converted LED.** **a,** Schematic of NIR imaging with the as-fabricated broadband NIR LED and a NIR camera. **b, c, d,** Representative imaging under ambient visible illumination (top) and NIR-LED illumination (bottom), with NIR-camera images highlighting enhanced contrast and subsurface information. **e,** Schematic of the spectroscopic sensing platform (NIR LED + integrating sphere + spectrometer). **f, g,** Normalized transmitted spectra of ten solvents, showing solvent-specific modulation of the three fingerprint peaks. **h,** 3D ratiometric feature space constructed from $x = I_{1540}/I_{996}$, $y = I_{1540}/I_{1220}$, $z = I_{1220}/I_{1540}$ for solvent identification, leveraging internal normalization to suppress intensity drift and baseline/scattering variations. **i,** DMF/$H_2O$ mixture quantification: spectra (top) and linear dependence of

ratiometric variables on composition (bottom). **j,** Proof-of-concept freshness assessment: photographs (top) and time-dependent NIR spectral maps (bottom) of orange peel during thermal aging (80 °C), revealing systematic spectral evolution associated with dehydration.

Capitalizing on the three fingerprint NIR emission peaks (996, 1220, 1540 nm) of HEA-Yb-Tm-Er single crystals across their tunable 1000–1700 nm range, a ratiometric detection strategy for organic solvents is achieved. The differential response of these peaks to characteristic absorptions of functional groups (e.g., O─H, C─H, N─H) grants this method higher selectivity and robustness against interference than single-wavelength detection, ensuring more reliable quantitative analysis and a richer understanding of solvent properties. [55] The setup shown in Fig. 5e consists of an NIR LED source, an integrating sphere, and an optical spectrometer, used to characterize samples via transmission or reflection. Following excitation by the LED, NIR light interacts with the sample and is directed to the spectrometer for spectral acquisition, with data subsequently stored and processed on a computer. The absorption spectra of ten solvents-including $H_2O$, methanol (MeOH), ethanol (EtOH), hexane (n- hexane), proprylene glycol monomethyl ether (PG ME), acetic acid (HAc), octylamine ($C8\text{-}NH_2$), aniline ($Ph\text{-}NH_2$), *N*, *N*-Dimethylformamide (DMF), and tetrahydrofuran (THF) -are clearly resolved (Supplementary Fig. S41). We performed chemometric analysis based on the initial NIR spectra of HEA-Yb-Tm-Er and the solvent-transmitted NIR spectra. The intensity and peak position variations in the NIR spectra of the ten solvents were detected, enabling solvent discrimination. Three characteristic absorption bands are observed at 950–1000 nm, 1150–1250 nm, and 1400–1550 nm, corresponding to the overtones of O─H, N─H and C─H stretching modes, respectively (Figs. 5f-5g).[56, 57]

Moreover, the three characteristic NIR emission peaks of HEA-Yb-Tm-Er single crystals were used to define ratiometric variables ($x = I_{1540}/I_{996}$, $y = I_{1540}/I_{1220}$, $z = I_{1220}/I_{1540}$), enabling quantitative comparison across solvents with different functional groups. As plotted in Fig. 5f, these three intensity ratios map each solvent to a distinct coordinate in a three-dimensional space. In this ratiometric coordinate system, solvents are spatially separated according to their unique spectral-response patterns, which significantly enhances identification accuracy of solvents and provides a clear visual basis for discrimination. Importantly, using three independent ratiometric variables provides built-in redundancy and cross-validation, which suppresses intensity fluctuation artifacts (e.g., excitation power, optical alignment, and sample loading) and thereby improves identification accuracy. A ratiometric approach was further applied to discriminate binary solvent mixtures, using $H_2O$-DMF as a model system. The corresponding spectral changes are displayed in Fig. 5i (top), where the intensities of the three NIR peaks (996, 1220, and 1540 nm) decrease systematically with increasing DMF content. This trend can be quantified via three intensity ratios, which all show strong linear correlations as the $H_2O$ concentration rises from 20 vol% to 100 vol%. The obtained correlation coefficients ($R^2$) of 0.9964, 0.9971, and 0.9980 confirm the reliability and quantitative consistency of the ratiometric detection system (Fig. 5i, bottom).

Building on the ratiometric multi-peak approach developed for mixed-solvent discrimination, we further extended its application to the quantitative assessment of freshness in solid biological samples. The NIR spectral responses clearly differentiated orange peel subjected to different drying treatments (fresh, 0.5 h at 80 °C, and 1 h at 80 °C), as shown in Fig. 5j. The systematic spectral variations—such as intensity changes or peak shifts in regions associated with O–H and C–H vibrations—directly correlate with the loss of moisture and the resulting physical withering. In summary, we demonstrate that configurational-entropy engineering can translate compositional complexity into a device-relevant advantage: multi-peak, self-referenced NIR output with improved environmental robustness. This platform and the mechanistic insights reported herein inform transferable design principles for engineering robust photonics in perovskite-related metal halides.

**Conclusion**

In summary, we establish a high-entropy rare-earth halide double-perovskite platform, $Cs_2Na(Sb,RE)Cl_6$, that converts configurational entropy from a phenomenological "stability label" into a designable handle that simultaneously delivers expanded NIR photonic functionality and mechanistically accountable environmental robustness. By near-equiatomic B(III)-site alloying ($\Delta S_{config} \approx 1.6R$), $Sb^{3+}$ is harnessed as a broadband sensitizer to unify excitation and cooperatively activate multiple lanthanide emission channels, transforming the parent single-mode response into a wide-coverage NIR output spanning ~850–1600 nm with three spectrally orthogonal fingerprint bands at 996, 1220, and 1540 nm. This tri-peak, self-referenced signature provides an intrinsically drift-resilient readout, enabling redundancy-based ratiometric solvent identification and quantitative mixture sensing, and it translates directly to spectrally stable phosphor-converted NIR LEDs. Beyond functional expansion, accelerated aging experiments demonstrate markedly improved phase and emission stability under extreme humidity and oxygen compared with single-component analogues.

Importantly, theory disentangles the origin of this robustness and resolves a key ambiguity in high-entropy perovskites: the stability gain does not primarily arise from weakened $H_2O/O_2$ adsorption, which remains comparable to the parent phases, but instead from a kinetic bottleneck. Molecular dynamics reveals strongly suppressed self-diffusion of both $RE^{3+}$ cations and $Cl^-$ anions in the high-entropy lattice, which kinetically impedes ion-migration-assisted compositional redistribution, lattice reconstruction, and the propagation of surface-initiated reactions into the bulk. Together with the $-T\Delta S_{config}$ stabilization term, this entropy-enabled transport suppression emerges as a quantitative, composition-transferable descriptor linking configurational entropy to operational durability. These results position high-entropy metal halides as more than phase-stable materials: they can be engineered into multi-channel, self-referenced photonic solids whose output fidelity is protected by a mechanistic rule rooted in inhibited ionic transport. Looking forward, extending this framework to other halide lattices, facets, and dopant manifolds—and coupling it with device-relevant stressors such as thermal cycling, optical flux, and interfacial electric fields—should enable predictive, descriptor-driven design of robust broadband emitters and sensors for next-generation photonics.

## Experimental Section

### Materials

Cesium chloride (CsCl, 99.99%), sodium chloride (NaCl, 99.99%), antimony chloride hexahydrate ($SbCl_3 \cdot 6H_2O$, 99.99%), erbium chloride hexahydrate ($ErCl_3 \cdot 6H_2O$, 99.99%), ytterbium chloride hexahydrate ($YbCl_3 \cdot 6H_2O$, 99.99%), thulium chloride hexahydrate ($TmCl_3 \cdot 6H_2O$, 99.99%), yttrium chloride hexahydrate ($YCl_3 \cdot 6H_2O$, 99.99%), and gadolinium chloride hexahydrate ($GdCl_3 \cdot 6H_2O$, 99.95%) were purchased from Aladdin. Scandium chloride hexahydrate ($ScCl_3 \cdot 6H_2O$, 99.99%) was purchased from Macklin Biochemical Co., Ltd. Hydrochloric acid (HCl, 37wt% in water) were purchased from Sinopharm Chemical Reagent Co., Ltd, China. All chemicals were used directly without further purification.

### Synthesis of $Cs_2NaRECl_6$ ($RE^{3+}$=$Er^{3+}$, $Yb^{3+}$, $Tm^{3+}$, $Sc^{3+}$) single crystals

The $Cs_2NaRECl_6$ single crystals were synthesized *via* the hydrothermal method following established procedures. 2 mmol CsCl, 1 mmol NaCl, 1 mmol $RECl_3 \cdot 6H_2O$ were dissolved in 4 mL HCl solvent (12 M) in a 25 mL Teflon-lined container. The mixture was subjected to heating at 180 °C for 12 hours using a stainless-steel Parr autoclave. And then, the reactor was cooled to the 25°C within 30 h. The resulting single crystals rinsed with anhydrous ethanol, and finally dried in a drying oven at 60 °C for 2h.

**Synthesis of $Cs_2Na(Sb_{0.2}Sc_{0.2}Er_{0.2}Yb_{0.2}Tm_{0.2})Cl_6$ single crystals (HEA-Yb-Tm-Er)**

The synthesis procedure for HEA-Yb-Tm-Er single crystals closely affinitive that of $Cs_2NaRECl_6$ single crystals. In a nutshell, 2 mmol CsCl, 1 mmol NaCl, 0.2 mmol $SbCl_3·6H_2O$, 0.2 mmol $YbCl_3·6H_2O$, 0.2 mmol $ErCl_3·6H_2O$, 0.2 mmol $TmCl_3·6H_2O$ and 0.2 mmol $ScCl_3·6H_2O$ was blended within a 4 mL HCl solution housed in a 20 mL stainless-steel Parr autoclave. The resulting mixture solution was subjected to a temperature of 180 °C for 12 hours, and then was slowly cooled to 25°C within 30 h.

**Synthesis of $Cs_2Na(Sb_{0.2}Sc_{0.2}Y_{0.2}Gd_{0.2}RE_{0.2})Cl_6$ single crystals (HEA-RE) ($RE^{3+}$= $Er^{3+}$, $Yb^{3+}$, $Tm^{3+}$)**

2 mmol CsCl, 1 mmol NaCl, 0.2 mmol $SbCl_3·6H_2O$, 0.2 mmol $YCl_3·6H_2O$, 0.2 mmol $GdCl_3·6H_2O$, 0.2 mmol $TmCl_3·6H_2O$ and 0.2 mmol $ScCl_3·6H_2O$ was blended within a 4 mL HCl solution housed in a 20 mL stainless-steel Parr autoclave. The resulting mixture solution was subjected to a temperature of 180 °C for 12 hours, and then was slowly cooled to 25°C within 30 h. The synthesis of HEA-Yb and HEA-Er single crystals are performed under otherwise identical conditions, but replaced the corresponding $RECl_3·6H_2O$.

**Characterization**

XRD measurements were performed on a SmartLab-SE X-Ray diffractometer (Cu Kα, λ = 1.5406 Å). SEM images and EDS mapping were captured using a field-emission scanning electron microscope (Gemini SEM 300) equipped with an Energy Dispersive Spectrometer (EDS). The selected-area electron diffraction (SAED) and high-resolution transmission electron microscopes (HRTEM) images were obtained with a JEOL JEM-ARM200F transmission electron micro scope operating at an acceleration voltage of 200 kV. X-ray photoelectron spectroscopy (XPS) data were collected using a Thermo SCIENTIFIC ESCALAB 250Xi spectrometer equipped with an Al Kα (1486.6 eV) line as excitation source. Inductively coupled plasma mass-spectrometry (ICP-MS) was performed on a Thermo iCAP Qc. Absorption spectra were collected using a Shimadzu UV-3600 UV/Vis/NIR spectrophotometer. Photoluminescence (PL) and photoluminescence excitation (PLE) spectra were recorded by a QM8000 HORIBA spectrometer. Notably, the long pass filter of 850 nm should be applied during the measurement of NIR spectra. The PL decay curves were obtained by a time-correlated single-photon counting (TCSPC) lifetime spectroscopy system with a SpectraLED (340 nm) as the light source. Photographs for the NIR LEDs were captured by taken by an industrial night-vision camera (MVCA050-20GN, Hikvision, China). The electroluminescence (EL) performance was measured by a fiber spectrophotometer (NIR17S spectrometer, Idea Optics, China) in the spectral region from 850 nm to 1600 nm.

**Preparation of LED devices**

NIR LEDs were fabricated by integrating UV LED chips ($\lambda_{ex}$ = 340 nm, Shenzhen Fangpu photoelectric Co., LTD, China; FOVNNBFZ1.8W320) with HEA-Yb-Tm-Er perovskite materials. Typically, epoxy resin A and B were weighed with a ratio of 4: 1, and then perovskite material was thoroughly mixed with epoxy resin. The mixture was then spread on surface of the UV chips and transferred in the oven followed by drying at 100 °C for 40 min to produce the NIR LED devices. The electroluminescence performance of NIR-LEDs was operated under a voltage of 3 V and currents at 100 mA.

**Computational Methodology**

**Electronic structure calculations**

All density functional theory (DFT) calculations were performed using the Vienna Ab initio Simulation Package (VASP).[58] The projector augmented-wave (PAW) pseudopotentials were used to treat the electron-ion interactions.[59] The generalized gradient approximation (GGA) in the Perdew–Burke–Ernzerhof (PBE) form was employed as the exchange-correlation functional to handle the interactions between electrons.[60] The unit cell of the pure-phase double perovskite studied in this work contains 40 atoms, and a 2×2×2 supercell of 320 atoms was constructed to obtain a high-entropy perovskite structure with random distribution of rare-earth elements in a sublattice. Due to the inherent limitations of conventional Density Functional Theory (DFT) in dealing with localized strongly correlated electrons (e.g., the 4f orbitals of rare-earth elements), in this work, we applied the Hubbard U correction proposed by Dudarev et al.[61] to the f localized orbitals of rare-earth elements (Er, Tm, Yb) based on conventional DFT functionals, and considered on-site Coulomb interactions on the d orbitals of Er, Tm, and Yb with effective values of U = 6.5, 7.0, 6.5 eV and J = 0.8, 0.9, 0.8 eV, respectively. This introduces electron correlation energy, compensates for the aforementioned shortcomings, and makes the computational results more consistent with experiments. The plane-wave cutoff energy was set to 342 eV. The convergence criteria were set to $10^{-5}$ eV for the total energy and 0.01 eV Å−1 for the force, respectively. For pure-phase double perovskites, a 3 × 3 × 3 Γ-centered Monkhorst−Pack k-point grid was used to sample the Brillouin zone in both geometry optimization and electronic structure calculations (including band structure, total and projected density of states). For high-entropy perovskites, due to the large supercell, only Γ point was used to sample the Brillouin zone in both geometry optimization and electronic structure calculations. The hybrid functional of Heyd−Scuseria−Ernzerhof (HSE06) was employed to obtain more accurate bandgaps based on the PBE relaxed structures.[62]

**Calculation of adsorption energy**

To calculate the adsorption energies of $O_2$ and $H_2O$ on the surface of perovskites, the slab models of (001) facet were constructed based on the 320-atom supercells of the VASP relaxed equilibrium structures of the HEA-Yb-Tm-Er, $Cs_2NaYbCl_6$, $Cs_2NaTmCl_6$, and $Cs_2NaErCl_6$. A vacuum space of 15 Å was applied to minimize the interactions between periodic images. The adsorption energy $E_{ads}$ was calculated according to the following eq 2:

$$E_{ads} = E_{(sub+molecule)} - (E_{sub} + E_{molecule}) \quad (2)$$

**Molecular dynamics simulations**

All the molecular dynamics simulations were performed on the equilibrium structures obtained from fully relaxed structures by using VASP. Molecular dynamics simulations were tested, simulated, cross checked by the LASP package[63] with the G-NN potential[64] and the LAMMPS package[65] with the DPA-3.1-3M potential.[66] The main calculations in this study were performed via LASP with cross check by LAMMPS. To evaluate the stability of the structure in water-oxygen environment, MD simulations in the canonical (NVT) ensemble were conducted at 500 K for a duration of 50 ps. The surface model of MD simulation was constructed as a six-layer (001) slab with a 20 Å vacuum space applied to both the upper and lower surfaces to eliminate periodic interactions. The ambient environment was simulated by introducing 40 $H_2O$ molecules and 20 $O_2$ atoms into the vacuum space. The mean squared displacement (MSD) of atoms was calculated from the NVT trajectories via eq 3,

$$\mathrm{MSD} = \frac{1}{N}\sum_{i=1}^{N}[R_i(t+t_0) - R_i(t_0)]^2 \quad (3)$$

where N is the total number of mobile ions and $R_i(t+t_0)$ and $R_i(t_0)$ are the positions of atom $i$ at time $t+t_0$ and $t_0$, respectively.

The corresponding diffusion coefficients (D) were extracted from the time dependence of the MSD via eq 4.

$$\mathrm{D} = \lim_{t\to\infty}\frac{R^2(t)}{2dt} \quad (4)$$

where d represents the dimensionality.

## Acknowledgements

This work was financially supported by the National Key R&D Program of China (Grant Number 2023YFB3608903), the National Natural Science Foundation of China (22305089, 22575093), Guangdong Provincial Basic and Applied Basic Research Fund (2025-2027) and the Guangdong Basic and Applied Basic Research Foundation (2025A1515011642). This work was also supported by the Interdisciplinary Research Support Program (2025JCYJ010) of the HUST Independent Innovation Program and the Open Topics Program (2025-6) funded by Hubei Key Laboratory of Bioinorganic Chemistry and Materia Medica. The authors acknowledge Dr. Xiaoli Gao of the Chemical Experimental Teaching Center, School of Chemistry and Chemical Engineering, Huazhong University of Science and Technology for her assistance with morphology measurement by Scanning

Electron Microscopy (SEM Hitachi SU8010). We acknowledge the Analytic & Testing Center (HUST) for Materials Characterizations. The computation is completed in the HPC Platform of Huazhong University of Science and Technology.

**Author contributions**

Y.X., L. M. Y., and Z.Z. conceived the idea and designed the experiments. L. M. Y., and Z.Z. led and coordinated the overall project. Y.X., J.W., X.C. and Y.Y. carried out experiments and characterizations. S.Y., L.L., F.Y. and J.T. performed the HRTEM measurements. C.Y. and L. M. Y. elucidated the theoretical insights of mechanism, performed DFT calculations and molecular dynamics simulations with the help and support from Y. W., H. W., G. C., Z. L.. J. J. and M. L. contribute to the discussion in the experiment and data analysis. Y.X. and C.Y. prepared the paper. Z.Z. and L. M. Y. revised the paper. All the authors contribute to the discussion and finalization of this work. Y.X. and C.Y. contributed equally to this work.

## Conflict of Interest

The authors declare no conflict of interest.

## Data Availability Statement

The data that support the findings of this study are available from the corresponding author upon reasonable request.